\documentclass{article}
\usepackage{spconf,amsmath,graphicx, booktabs, float}
\usepackage{multirow, tabularray, hyperref, setspace, enumerate,enumitem}

\title{Unified Pretraining Target Based Video-Music Retrieval with\\Music Rhythm and Video Optical Flow Information}

\name{Tianjun Mao$^1$ , Shansong Liu$^2$$^\dagger$\thanks{This work is done when Tianjun Mao is an intern at Tencent PCG ARC lab. Shansong Liu is the corresponding author. The first two authors contribute equally.}, Yunxuan Zhang$^3$, Dian Li$^3$, Ying Shan$^2$}
\address{ $^1$Fudan University \\
  $^2$ARC Lab, Tencent PCG \\
  $^3$ Fundamental Technology Center, Tencent PCG}

\begin{document}

\ninept
\maketitle
\begin{abstract}
Background music (BGM) can enhance the video's emotion. However, selecting an appropriate BGM often requires domain knowledge. This has led to the development of video-music retrieval techniques. Most existing approaches utilize pretrained video/music feature extractors trained with different target sets to obtain average video/music-level embeddings. The drawbacks are two-fold. One is that different target sets for video/music pretraining may cause the generated embeddings difficult to match. The second is that the underlying temporal correlation between video and music is ignored. In this paper, our proposed approach leverages a unified target set to perform video/music pretraining and produces clip-level embeddings to preserve temporal information. The downstream cross-modal matching is based on the clip-level features with embedded music rhythm and optical flow information. Experiments demonstrate that our proposed method can achieve superior performance over the state-of-the-art methods by a significant margin.
\end{abstract}
\section{Introduction}
\label{sec:intro}

\noindent
The recent explosion of short videos on various platforms across the world has attracted great attention and significantly raised the need for video-music retrieval. Video creators frequently use background music (BGM) in their posted short videos, which can help creators better convey their emotions. However, manually choosing an appropriate BGM for a video often requires domain knowledge or a thorough comprehension of the video. Therefore, using methods like cross-modal matching based video-music retrieval to recommend BGM for videos has attracted lots of interest.

Some pioneering methods explored the use of metadata \cite{brochu2003sound,mayer2011analysing,chao2011tunesensor,acar2014understanding,wu2016bridging,liu2018background} to conduct cross-modal matching based music recommendation. Chao \textit{et al.} \cite{chao2011tunesensor} calculated the correlation between photo album tags and music emotion tags to perform music recommendations for photo albums. Acar \textit{et al.} \cite{acar2014understanding} and Liu \textit{et al.} \cite{liu2018background} relied on Thayer’s Valence-Arousal model \cite{thayer1990biopsychology} to classify the video-music clips into the emotion quadrant to which they may belong. In recent years, Hong \textit{et al.} \cite{hong2018cbvmr} proposed a content-based audio-visual network for video-music cross-modal matching. Li and Kumar \cite{li2019query} utilized different emotion tag sets to pretrain video and music feature extractors for downstream video-music alignment. Apart from matching audio-visual modalities only, Yi \textit{et al.} \cite{yi2021cross} employed the textual modality as an additional input which is the music metadata in the video stream to obtain fused video-text embeddings before aligning with the music modality. In \cite{pretet2022video}, Pr{\'e}tet \textit{et al.} proposed a segment-level audio-visual cross-modal matching approach for video-music retrieval, still, temporal hints like music rhythm and video optical flow information have not yet been taken into consideration.

Most of these existing video-music retrieval approaches adopted pretrained networks trained with different target sets \cite{hong2018cbvmr,li2019query,yi2021cross,pretet2021cross} to produce average video/music-level embeddings. The drawbacks are two-fold. One is that different target sets for video/music pretraining may cause the generated embeddings difficult to match. For instance, Pr{\'e}tet \textit{et al.} \cite{pretet2021cross} used various audio pretraining networks as music feature extractors.
Among them, the worst-performing audio-pretrained network has a recall rate up to 13\% lower than the best network. The other one is using average embeddings neglects the potential temporal correlation between video and music.

For the first issue, we propose an approach that leverages a unified target set, which is produced by a tagging model followed by a TF-IDF \cite{luhn1957statistical,jones1972statistical} calculation process (see Sec. \ref{sec:unitag} for more details), to perform video/music pretraining. The pretrained models output clip-level video/music embeddings to preserve temporal information. We combine these clip-level embeddings from the same music or video in sequential order before feeding them into the following Transformer \cite{vaswani2017attention} encoders. For the second issue, the downstream cross-modal matching model is equipped with a cross-modal attention mechanism to fuse the video and music modalities. Inspired by \cite{mercea2022audio}, we treat the target labels as an additional textual modality to facilitate video-music alignment. What is more, both the video frames and audio contain rhythm information. For example, variations in the movement speed of people or objects in the video, or the alternations of light and heavy beats in the audio, can provide such information. When people search for background music for videos, they often tend to choose music that has similar rhythmic characteristics to the video. Thus, we adopt music rhythm and video optical flow information as auxiliary temporal hints to instruct the video-music cross-modal matching.

The contributions of our work are three-fold:
\begin{itemize}[itemsep=2pt,topsep=0pt,parsep=0pt]
\item Presenting the use of a unified target label set to pretrain video/music feature extractors for downstream video-music cross-modal matching.
\item Exploring the impact of temporal information including embedded rhythm and optical flow features.
\item Demonstrating superior performance over the state-of-the-art video-music retrieval methods. Demo and code could be seen in \href{https://melissamao.github.io/ut_cmvmr.github.io/}{https://melissamao.github.io/ut\_cmvmr.github.io/}.
\end{itemize}

The rest of this paper is organized as follows. Section \ref{sec:unitag} describes the acquisition of the unified target label set and video/music pretraining as well as details of the video-music cross-modal matching framework. The experiments are described in Section \ref{sec:exp}. We conclude our work in Section \ref{sec:conclusion}.

\section{Model Architecture}
\label{sec:unitag}
\subsection{Unified Target Label Set}

A unified target label set is used for video and music pretraining. Existing approaches \cite{hong2018cbvmr, li2019query, yi2021cross, pretet2021cross, li2021deep} used different target sets to train feature extractors for different modalities, which can lead to matching problems. For the video modality, many pretrained models are usually based on image classification tasks, while for the music modality, the objectives of pretraining models are typically audio classification \cite{gemmeke2017audio} or music tagging \cite{pons2019musicnn}. As described in \cite{pretet2021cross}, the best performed audio pretrained model OpenL3 \cite{cramer2019look,arandjelovic2017look} was trained towards an audio-visual correspondence objective, while the MuSimNet \cite{pretet2020learning} was trained on a music similarity task, demonstrating a much lower performance compared to the OpenL3 model. This to some extent shows that different target sets or objectives for video/music pretraining may cause the generated embeddings difficult to match. The unified target set establishes a unified objective for both video and music pretraining, enhancing compatibility between the modalities.

On popular short video platforms like TikTok or Kuai-shou, hashtags are commonly used with videos, they can provide meta-information for videos and facilitate user's video search. In the following description, hashtags will be referred to as tags. We employ an internal video multi-tagging model\footnote{The video multi-tagging model is not a building block of our proposed model. We just utilize it to obtain the video tags.} to generate tags for videos. These tags are then mapped to videos' corresponding background music, aiding in connecting video and music. In our data collection, one piece of music may correspond to multiple videos, and one video has several tags produced by the multi-tagging model. We utilize the TF-IDF algorithm \cite{luhn1957statistical,jones1972statistical} to simplify the target label set for pretraining video and music feature extractors. For each music $m$, we obtain a collection of video tags, which are not deduplicated. For a specific piece of music $m_i$, we can calculate the TF-IDF value of the $j$-$th$ tag of its corresponding tag collection $tagcol_i$ as follows:

\vspace{-0.2cm}
\begin{equation}
    \begin{aligned}
        TF &= \frac{M^j_{tagcol_i}}{M_{tagcol_i}} \\
        IDF &= \log\frac{N}{N^j_{tagcol_i}+1}
    \end{aligned}
\end{equation}

\noindent
where $M_{tagcol_i}$ is the total number of tags without deduplication in the tag collection $tagcol_i$ of music $m_i$, $M^j_{tagcol_i}$ denotes the number of times that the $j$-$th$ tag appears in $tagcol_i$, $N$ is the total number of music, and $N^j_{tagcol_i}$ indicates the number of music containing the $j$-$th$ tag of $tagcol_i$. 

\begin{figure*}
 \vspace{-0.5em}
    \centering
    \includegraphics[width=15.5cm]{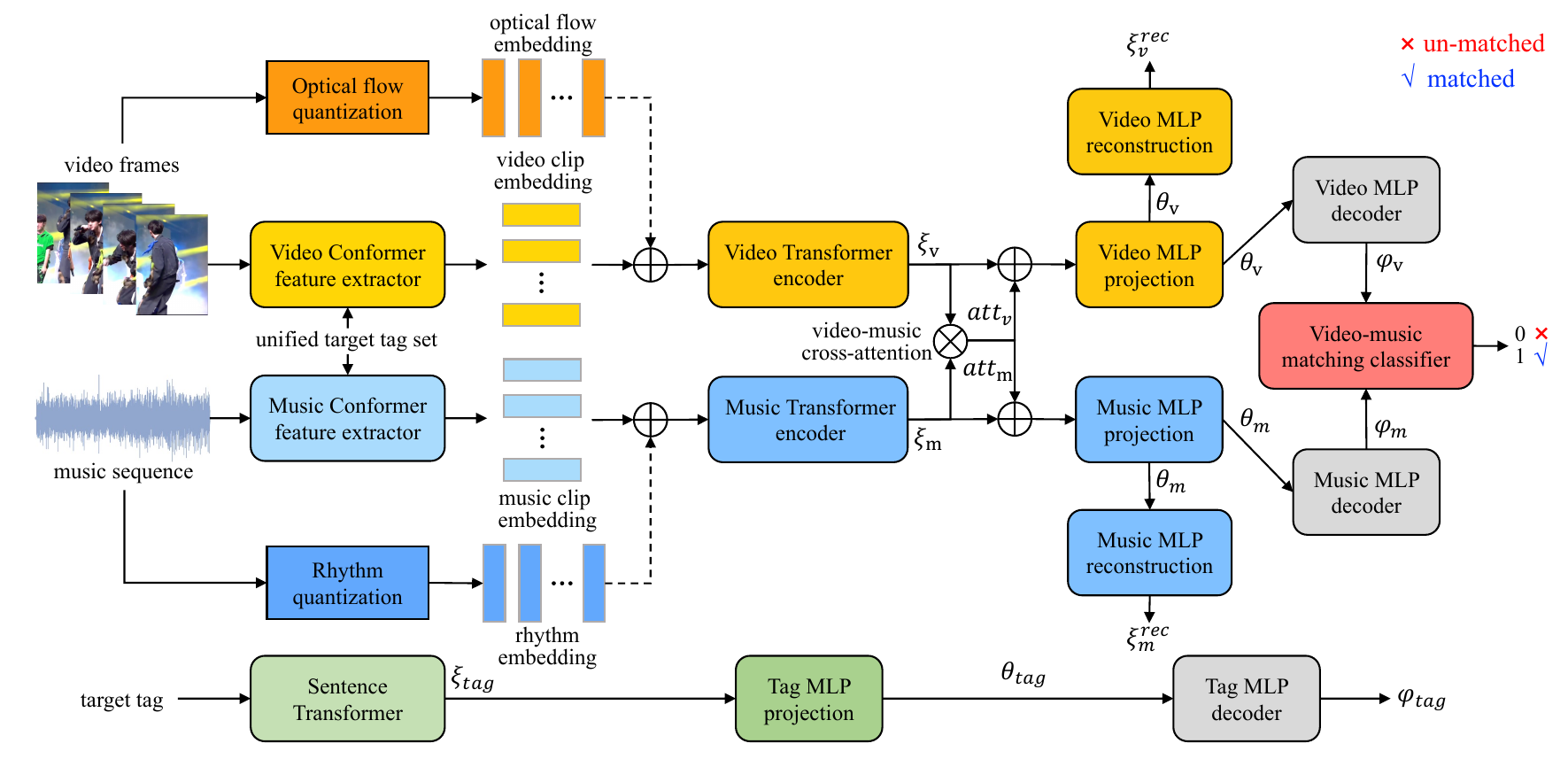}
    \caption{Illustration of the unified pretraining target based cross-modal video-music retrieval (UT-CMVMR) framework.}
    \label{fig:framwork}
     \vspace{-0.5em}
\end{figure*}

Using the equation above, TF-IDF values are calculated for each tag in a music's tag collection. The tag with the highest TF-IDF value becomes the target label for that music. The video label is assigned based on the video-music pairing. This process results in a unified target label set having 578 unique tags for video and music pretraining. As depicted in Fig. \ref{fig:framwork}, in the video/music pretraining block, the training objective is based on the unified target tags for both video and music Conformer \cite{gulati2020conformer} feature extractors. Full-length videos and music are trimmed to the same length. We denote a length-normalized video-music pair as $(\mathbf{V};\mathbf{M})$. To obtain clip-level video/music embeddings containing temporal information, the length-normalized videos and music are further chopped into clips, denoted as $(\mathbf{v_1,v_2,...,v_T};\mathbf{m_1,m_2, ...,m_T})$, where $T$ is the number of clips of a given $(\mathbf{V};\mathbf{M})$. Afterwards, we can use the pretrained video/music Conformers to extract clip-level embeddings $(\mathbf{f_{v_1},f_{v_2},...,f_{v_T}};\mathbf{f_{m_1},f_{m_2},...,f_{m_T}})$.



\subsection{Cross-modal Video-music Retrieval}
\label{sec:crossret}

The proposed unified pretraining target based cross-modal video-music retrieval (UT-CMVMR) framework is illustrated in Fig. \ref{fig:framwork}, consisting of two main branches, which are video and music streams, and an auxiliary textual branch. It is worth noting that the textual branch based on tags is an optional component during the training of the downstream video-music matching task. Its purpose is to further strengthen the connection between video and music.

\noindent
\textbf{1) Optical Flow and Rhythm Quantization:}
Rhythmic characteristics are crucial for video background music selection. Inspired by \cite{di2021video}, music videos often have background music that matches the rhythm of the visual content. For example, fast-paced moving visuals in a video (such as sports videos) are suitable for pairing with a lively and fast-paced musical piece, while tranquil café scenes are better suited for pairing with a soothing music genre like jazz. In this work, regarding video-music rhythmic consistency, we adopt music rhythm and optical flow information to better leverage the temporal relation between video and music. We define the clip-level music rhythm information by three statistics, consisting of the number of beats in a clip ($N_{beat}$), the average beat strength of a clip ($S_{beat}$), and the average interval length between two consecutive beats in a clip ($\overline{L}$). The video optical flow information is calculated by the average pixel displacement between adjacent frames in a clip ($\overline{M}$). Except for the discrete $N_{beat}$, which can be counted directly, the other three statistics $S_{beat}$, $\overline{L}$ and $\overline{M}$ are continuous values, so we estimate their range and utilize bins to discretize them. The $nn.Embedding$ in Pytorch \cite{paszke2019pytorch} is used to obtain feature embeddings for the discretized rhythm and optical flow information. Afterwards, the clip-level rhythm and optical flow embeddings can be regarded as additional plug-in inputs to video and music streams of the UT-CMVMR framework shown in Fig. \ref{fig:framwork}.

\noindent
\textbf{2) Downstream Video-Music Matching:}
The input of the subsequent video-music matching module is clip-level embeddings, passing through two different Transformer encoders $V_e$ and $M_e$,

\begin{equation}
    \begin{aligned}
        \xi_{v} &= V_e(f_{v_1},f_{v_2},...,f_{v_T}) \\
        \xi_{m} &= M_e(f_{m_1},f_{m_2},...,f_{m_T}) 
    \end{aligned}
\end{equation}

\noindent
where $\xi_{v}$ and $\xi_{m}$ are the embeddings generated by the video and music encoders. The downstream structure of our proposed framework is depicted in Fig. \ref{fig:framwork}, in which a video-music cross-attention mechanism is used to fuse video and music modalities. With the assistance of this module, the embedding spaces of music and video are brought closer, enhancing the ability to match information from these two modalities. The outputs of the cross-attention block are $att_v$ and $att_m$. The calculated attention is then summed with the embeddings from the previous video/music encoder for the following multilayer perceptron (MLP) projection,

\vspace{-0.1cm}
\begin{equation}
    \begin{aligned}
        \theta_{v} &= V_p(\xi_{v} + att_v) \\
        \theta_{m} &= M_p(\xi_{m} + att_m) 
    \end{aligned}
\end{equation}

\noindent
where $\theta_{v}$ and $\theta_{m}$ are the video/music embeddings after video-music cross-attention and linear projection. The video/music MLP reconstruction and decoder blocks serve as the regularization components to constrain the model training. The last video-music matching classifier is to determine whether a video-music input sample is matched or not. For the textual modality, the input is the common tag label of a video-music pair. Its function is to bring the music and video representations closer during training while distinguishing video representations corresponding to the same music. The tags are passed by a sentence Transformer \cite{reimers-2019-sentence-bert} followed by an MLP projection to obtain the textual embedding $\theta_{tag}$. The weights of the last MLP decoder of the textual modality are shared with the video/music MLP decoders. 

\vspace{-0.3em}
\subsection{Loss Functions}
\label{subsec:loss}
\noindent
\textbf{1) Triplet loss:} We train our UT-CMVMR model mainly by several triplet loss functions. The basic definition of a triplet loss is $t(x^+,y^+,y^-)=max(d(x^+,y^+)-d(x^+,y^-)+\lambda,0)$, where $x^+$, $y^+$ and $y^-$ denote anchor, positive and negative, respectively. $d(\cdot,\cdot)$ means the Euclidean distance, and $\lambda$ is the margin. 
Inspired by \cite{mercea2022audio}, we form three types of triplet losses to ensure the projected information of three modalities is aligned with the guidance of tags, which are written as follows,
\vspace{-0.1em}
\begin{equation}
    \begin{aligned}
        \mathcal{L}_{av} &= t(\theta_v^+,\theta_m^+,\theta_m^-) + t(\theta_m^+,\theta_v^+,\theta_v^-) \\
        \mathcal{L}_{vtag} &= t(\theta_v^+,\theta_{tag}^+,\theta_v^-) + 
        t(\theta_{tag}^+,\theta_{v}^+,\theta_{tag}^-) +
        t(\varphi_{tag}^+,\varphi_{v}^+,\varphi_v^-)
        \\
        \mathcal{L}_{atag} &= t(\theta_m^+,\theta_{tag}^+,\theta_m^-) + 
        t(\theta_{tag}^+,\theta_m^+,\theta_m^-) +
        t(\varphi_{tag}^+,\varphi_{m}^+,\varphi_m^-) 
    \end{aligned}
\end{equation}

\noindent
\textbf{2) Regularization loss:} The purpose of employing regularization loss is to use tag embeddings as a bridge to bring the embedding spaces of audio and video closer while preserving the distinctive information of their respective modalities. The mean squared error (MSE) function is used in the regularization loss, including three groups, which are $d(\xi_v,\xi_v^{rec})+d(\xi_m,\xi_m^{rec})$, $d(\varphi_v,\xi_{tag})+d(\varphi_m,\xi_{tag})+d(\varphi_{tag},\xi_{tag})$ and $d(\theta_v,\theta_{tag})+d(\theta_m,\theta_{tag})$. Therefore, the regularization loss can be formulated as follows,

\vspace{-0.3cm}
\begin{equation}
    \begin{split}
        \mathcal{L}_{regular} &= d(\xi_v,\xi_v^{rec}) + d(\xi_m,\xi_m^{rec}) + d(\varphi_v,\xi_{tag}) \\
        &+ d(\varphi_m,\xi_{tag}) + d(\varphi_{tag},\xi_{tag}) + d(\theta_v,\theta_{tag}) \\
        &+ d(\theta_m,\theta_{tag})
    \end{split}
\end{equation}

\noindent
\textbf{3) Cross-entropy matching loss:} In the video-music matching classifier shown in Fig. \ref{fig:framwork}, we adopt the cross-entropy loss $\mathcal{L}_{\text {ce}}$ for an additional binary classification task, which is to determine whether an input is a valid video-music pair or not.

\noindent
\textbf{4) Final loss function:} In training stage, we obtain the loss function $\mathcal{L}=\lambda _{1} \mathcal{L}_{\text {av}}+\lambda _{2} \mathcal{L}_{\text {vtag}}+\lambda _{3} \mathcal{L}_{\text{atag}}+\lambda _{4} \mathcal{L}_{\text {regular}}+\lambda _{5} \mathcal{L}_{\text {ce}}$.

\vspace{-0.2cm}

\vspace{0.3cm}
\section{Experiments}
\label{sec:exp}
\subsection{Data Preprocessing}
\vspace{-0.1cm}

Each video is decoded at 1 frame per second and each frame is converted to grayscale and resized to $224\times224$ pixels. Music is resampled to 16kHz. We trim the videos and music to the same length (no more than 28 seconds) and further chop them into 4-second clips. The video/music clips shorter than 4 seconds are padded to 4 seconds, where video clips use last-frame padding and music clips adopt zero padding. For a video clip, we concatenate the frames of a clip by time to form an $896\times224$ matrix. For a music clip, we compute its filter-bank (FBK) features, where the number of Mel-frequency bins is set as 80, resulting in a $398\times80$ matrix. Then these preprocessed video/music clips are used to pretrain the video/music Conformers for clip-level embedding generation.

\vspace{-0.2cm}
\subsection{Implementation Details}
\vspace{-0.1cm}

\textbf{1) Pretraining models:} 
An open-source Conformer implementation \cite{kim2020conformer_pytorch} is used for video/music pretraining. The Conformer blocks are set to 6 and 2 for video and music. Both models share a feature embedding dimension of 512. Sentence-transformer \cite{reimers-2019-sentence-bert} is used for tag embedding\footnote{\url{https://www.sbert.net/}}, with an embedding dimension of 768.

\noindent
\textbf{2) Cross-modal matching model:} We implement our downstream video-music retrieval model referring to the AVCA framework proposed in \cite{mercea2022audio}. Video/music clips from the same video/music are first combined in sequential order and padded to $max$\_$seq$, where $max$\_$seq$ denotes the maximum number of clips of video/music sequences. Then they are fed into their respective video/music Transformer encoders. The dimensions of $\{\xi_v,\xi_m,\xi_{tag},\varphi_v,\varphi_m,\varphi_{tag},\\\xi_v^{rec},\xi_m^{rec}\}$ are set as 768, while the dimensions of $\{\theta_v,\theta_m,\theta_{tag}\}$ are set as 256 (see the notations in Fig. \ref{fig:framwork}).


\noindent
\textbf{3) Music rhythm and optical flow embedding:}We apply the $librosa$ toolkit \cite{mcfee2015librosa} to extract beat information $N_{beat}$, $S_{beat}$ and $\overline{L}$ from music clips. Their embedding dimensions are set to 256, 128, 128, respectively. Then we concatenate them to form a 512-dim embedding vector. A publicly available RAFT model \cite{teed2020raft} for optical flow extraction is used to obtain the clip-level average pixel displacement $\overline{M}$ of adjacent frames. The embedding dimension of $\overline{M}$ is set as 512 to match the music rhythm embedding dimension. The $nn.Embedding$ \cite{paszke2019pytorch} is adopted to convert the discretized rhythm and optical flow to continuous feature vectors.

\noindent
\textbf{4) Parameter settings:} The best hyper-parameters are determined by searching a combination from triplet margin \{0.01, 0.1, 0.5, 1, 2, 4\}, learning rate \{0.01, 0.001, 0.0001, 0.00001\} and weight \{0, 0.5, 1, 2\} of different loss functions (Sec. \ref{subsec:loss}).

\noindent
\textbf{5) Evaluation method:} The Recall@K is used to evaluate the performance of video-music retrieval systems following \cite{hong2018cbvmr}.

\vspace{-0.4cm}
\begin{table}[htbp]
\centering
\caption{Video-music retrieval performance of baseline CBVMR, CMVAE, CMVMR and our proposed UT-CMVMR systems on video-music pairs of our internal dataset.}
\vspace{0.2cm}
\label{tab:table}
\scalebox{0.98}{\begin{tabular}{c|c|c|c|c|c|c} 
\hline\hline
\multirow{2}*{Sys} & \multirow{2}{*}[-0.4ex]{Model} & \multirow{2}{*}[-0.4ex]{Setting} & \multicolumn{4}{c}{Recall@K (\%)}  
\\ 
\cline{4-7}
~ & ~ & ~ & K = 1  & K = 5   & K = 10  & K = 25   
\\ 
\hline
1 & CBVMR & \multirow{2}{*}[-0.4ex]{AE} & 4.54 & 15.69 & 27.71 & 43.35                   
\\
2 & CMVAE &   & 4.95    & 18.00     & 29.70     & 43.86                           
\\
\hline
3 & \multirow{4}{*}{CMVMR} & AE  & 5.13 & 19.83 & 30.32 & 44.80  \\                  
4 & ~ &A-SE & 5.18  & 22.38 & 35.76 & 45.50  
   \\
5 & ~ &SE  & 5.58 & 21.02 & 35.80 & 46.11   
 \\
6 & ~ & SE\&R  & \textbf{8.82} & \textbf{22.92} & \textbf{36.28} & \textbf{53.82} \\
\hline\hline
\end{tabular}}
\end{table}


\vspace{-0.3cm}
\subsection{Results on the Internal Dataset}
\vspace{-0.1cm}
\label{exp:main}
We conduct a set of experiments to verify the efficacy of our proposed UT-CMVMR framework. See results in Table. \ref{tab:table}. 

\noindent

\noindent
\textbf{1) The effectiveness of CMVMR structure with video-music cross-attention:} 
We conducted experiments using CBVMR, CMVAE, and CMVMR with average embeddings (AE) as in \cite{hong2018cbvmr}, where temporal information is ignored. CMVMR is the downstream cross-modal matching model in our UT-CMVMR framework. We use open-source pretrained EfficientNet-b5 and Musicnn models from \cite{Morris2022cross-retrieval_model} to generate video/music-level embeddings for these experiments. The results (Sys. 3 vs. Sys. 1-2) demonstrate that the CMVMR model outperforms other models in terms of all metrics.

\noindent
\textbf{2) The effectiveness of Conformer Extractor comes from both temporal information and the unified tag set:} 
The experiment (Sys. 4) with averaged clip-level features of a sequence (A-SE) is better than experiments with average embeddings (AE) extracted from open-sourced models \cite{Morris2022cross-retrieval_model} which is used in former three experiments (Sys. 1-3), especially when $K=5$ and $K=10$. This result is sufficient to prove that even if the timing information is ignored, the Conformer feature extractors trained with a uniform tag set can still perform better than the general audio/video feature extractor.

\noindent
\textbf{3) Effectiveness of temporal information and rhythm information:} Firstly, the pretrained video/music Conformers are used to generate clip-level embeddings (SE) to construct video-music feature sequences. Overall, compared to utilizing averaged clip-level features without temporal information, it works better (Sys. 5 vs. Sys. 4), despite the small degradation when $K=5$. Therefore, to better utilize the underlying temporal correlation between video and music, our final experiment (Sys. 6) is equipped with clip-level embeddings (SE\&R) as well as quantized music rhythm and optical flow information. It can be seen that, with music rhythm and optical flow embeddings, our final experiment using the proposed UT-CMVMR framework outperforms all the previous experiments. 

\vspace{-0.1cm}
\subsection{Results on HIMV-200K Dataset}

To rigorously illustrate the effectiveness of our proposed approach, we also test our approach on the public HIMV-200K dataset \cite{abu2016youtube}. 
For the data splitting of this dataset, we follow the settings of \cite{hong2018cbvmr}. As shown in Table. \ref{tab:table2}, we conduct experiments of the three models with settings of AE or SE\&R. The data preprocessing follow the experiments in Sec. \ref{exp:main}. From the experiments, we can observe that when using the same framework, the SE\&R based models significantly outperform those with AE, meaning that the temporal information plays an important role in video-music retrieval. 
The CMVMR framework consistently outperforms the baseline models, showcasing its generalization ability and effectiveness.

\vspace{-0.3cm}
\begin{table}[htbp]
\centering
\caption{Video-music retrieval performance of baseline CBVMR, CMVAE, CMVMR and UT-CMVMR systems on the HIMV-200K dataset with settings of AE and SE\&R.}
\label{tab:table2}
\vspace{0.2cm}
 \setlength{\tabcolsep}{0.8mm}{
\begin{tabular}{c|c|c|c|c|c}
\hline
\hline
\multirow{2}{*}[-0.4ex]{Model} & \multirow{2}{*}[-0.35ex]{Setting}                                   & \multicolumn{4}{c}{Recall@K(\%)}        
\\
\cline{3-6}

    \multirow{2}{*}{}  &          \multirow{2}{*}{}                                 & \multirow{1}{*}{K=1}          & \multirow{1}{*}{K=5}      & \multirow{1}{*}{K=10}   & \multirow{1}{*}{K=25}   \\
\hline

\multirow{2}{*}{CBVMR} & AE                             & 3.40          & 5.20  & 15.30  & 22.70 \\
      & SE\&R & 5.20          & 7.10  & 18.20  & 29.10  \\ 
\hline
\multirow{2}{*}{CMVAE} & AE                             & 4.70          & 9.10 & 17.00 & 41.20 \\
      & SE\&R & 6.10          & 11.80 & 20.40 & 44.00 \\ 
\hline
\multirow{2}{*}{CMVMR} & AE                             & 9.70          & 13.90 & 21.30 & 45.90 \\
      & SE\&R & \textbf{10.80}          & \textbf{28.10} & \textbf{36.50} & \textbf{51.60} \\
\hline
\hline

\end{tabular}}

\end{table}

\vspace{-0.3cm}
\subsection{Subjective Evaluation}

Apart from the objective evaluation using retrieval metrics to validate the efficacy of the proposed model, we conducted a subjective evaluation by inviting 24 people to compare the recommended music of the test videos from video-music pairs using our proposed and baseline models. 15 out of 24 people have knowledge of music theory or a basic understanding of music performance. Moreover, 21 out of 24 people maintain the habit of listening to music every week. Table. \ref{tab:table3} shows the percentages of each method they preferred. We can see that our method demonstrates the best performance. 

\vspace{-0.4cm}
 \setlength{\tabcolsep}{0.7mm}{
\begin{table}[htbp]
\centering
\caption{The preference result of the subjective evaluation.}
\label{tab:table3}
\vspace{0.2cm}
\begin{tabular}{c|c|c|c} 
\hline\hline
Model     & Our UT-CMVMR model & CBVMR & CMVAE  \\
\hline
Preferred & 50.00\%               & 40.00\%  & 10.00\%  \\
\hline\hline
\end{tabular}
\end{table}
\vspace{-0.2cm}

\section{Conclusion}
\label{sec:conclusion}

In this paper, we present the use of a unified target label set to pretrain video/music feature extractors for video-music cross-modal matching, providing ideas for processing and matching real-world audio-video data in industries. We also explore the use of temporal information for video-music retrieval. Experiments have demonstrated that even in the absence of a uniform tag set for the video-music matching task, adding temporal information still substantially improves the results, achieving superior performance over state-of-the-art methods. Possible future research will focus on further improving the video-music retrieval approach and its generalizability.

\clearpage


\bibliographystyle{IEEEtran}
\small\bibliography{refs}

\end{document}